\documentclass[%
 reprint,
nofootinbib,
 amsmath,amssymb,
 aps,
]{revtex4-1}

\usepackage{graphicx}
\usepackage{dcolumn}
\usepackage{bm}
\usepackage{upgreek}
\usepackage{sidecap}
\usepackage{natbib}
\usepackage{float}
\usepackage{blindtext}
\usepackage{color}

\begin{document}


\title{Interfacial flows in sessile evaporating droplets of mineral water}

\author{Massimiliano Rossi$^1$}
\email{rossi@fysik.dtu.dk}
\author{Alvaro Marin$^2$}%
\email{a.marin@utwente.nl}
\author{Christian J. K\"{a}hler$^1$}%
\affiliation{%
 $^1$Institute of Fluid Mechanics and Aerodynamics, Bundeswehr University Munich, 85577 Neubiberg, Germany.\\
 $^2$Physics of Fluids, University of Twente, 7522 NB Enschede,The Netherlands 
}%




\date{\today}

\begin{abstract}
Liquid flow in sessile evaporating droplets of ultrapure water typically results from two main contributions: a capillary 
flow \emph{pushing} the liquid towards the contact line from the bulk and a thermal Marangoni flow \emph{pulling} the drop free surface towards the summit. Current analytical and numerical models are in good qualitative agreement with experimental observations however they overestimate the interfacial velocity values by 2--3 orders of magnitude. This discrepancy is generally ascribed to contamination of the water samples with non-soluble surfactants, however an experimental confirmation of this assumption has not yet been provided. In this work, we show that a small ``ionic contamination'' can cause a significant effect in the flow pattern inside the droplet. To provide the proof, we compare the flow in evaporating droplets of ultrapure water with commercially available bottled water of different mineralization levels. Mineral waters are bottled at natural springs, are micro-biologically pure, and contain only traces of minerals (as well as traces of other possible contaminants), therefore one would expect a 
slower interfacial flow as the amount of ``contaminants'' increase. Surprisingly, our results show that the magnitude of the interfacial flow is practically the same for mineral waters with low content of minerals as that of ultrapure water. However, for waters with larger content of minerals, the interfacial flow tends to slow down due to the presence of ionic concentration gradients. 
Our results show a much more complex scenario than it has been typically suspected, and therefore a deeper and more comprehensive analysis of the huge differences between numerical models and experiments is necessary.

\end{abstract}

\maketitle

\begin{figure}[h!]
\centering
\includegraphics[width=.48\textwidth]{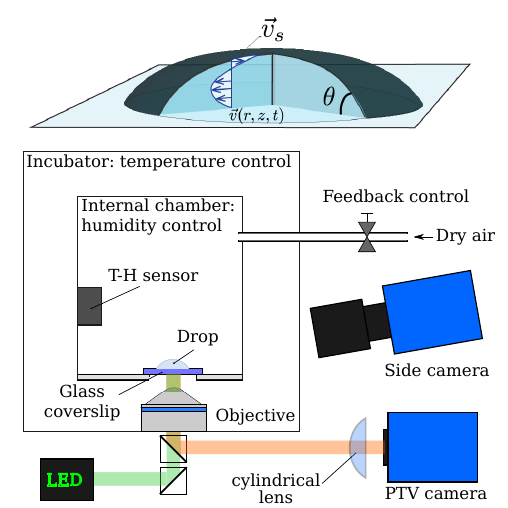} 
  \caption{Schematic of the experimental setup. The water droplet is deposited on a glass coverslip under constant temperature and humidity. A side view camera is used to measure the droplet shape. An epifluorescent microscope in combination with a cylindrical lens is used to image the tracer particles inside the droplets for the 3D-PTV method.} 
  \label{fig:sketch}
\end{figure}

Evaporating sessile droplets can be found in our everyday life when we spill coffee on our table \cite{deegan1997capillary}, they can be found in a booming industry as inkjet printing \cite{calvert2001inkjet,singh2010inkjet}, in crime scenes \cite{brutin2011blood,laan2015bloodstain}, and also in laboratories to perform analysis on their contents \cite{rios2018pattern,carreon2018texture}. 
When the droplets evaporate diffusively, simple analytical models can be used to predict their evaporation rate \cite{popov2005evaporative,gelderblom2011water} and consequently their evaporation time \cite{stauber2014lifetimes}, which is often of high interest in applications. 
However, the internal flows inside such droplets involve several complications. Although the capillary flow  described by Deegan \textit{et al.} \cite{deegan1997capillary,deegan2000contact} is typically the dominant one, evaporating sessile droplets are extremely sensitive systems, in particular regarding their interfacial flow: Small imbalances in the thermal equilibrium \cite{hu2005analysis,hu2006marangoni}, composition \cite{marin2016surfactant,marin2018solutal} or even the external environment \cite{malinowski2018dynamic}, can generate Marangoni flows that can overcome the bulk capillary flow.  
Water, also known as the ``universal solvent'' for being the most common polar liquid on Earth, is particularly complex to analyze due to its facility to dissolve all type of substances. While the response of the interfacial flow to surfactants is fairly well-known \cite{marin2016surfactant,still2012surfactant}, nothing has been reported for water solutions containing small amounts of mineral salts, very common in buffer solutions, biological samples or even in  water from faulty deionization systems that can be found in laboratories.  

An important reason for the lack of progress in the understanding of such interfacial flows is the fact that they occur along a moving curved surface and therefore they are complex to observe experimentally. Some efforts have been recently made either using fast confocal microscopy \cite{bodiguel2010imaging} or using optical coherence tomography \cite{edwards2018density}. Unfortunately, both lack a proper temporal resolution and are only able to track particles in short periods of time (optical coherence tomography) or scanning one plane at a time (confocal microscopy).

In this paper, we analyze the full 3D flow field (bulk and interfacial flow) of evaporating sessile droplets of ultrapure water (Milli-Q\textsuperscript{\textregistered}), typically used in laboratories, and we compare it with commercial bottled non-carbonated water from three european brands (ViO, Vittel and Gerolsteiner) with different mineralization levels. 
Our results show a significant impact in the flow when small amounts of dissolved ions are introduced in the solutions. Interestingly, once the liquid phase is evaporated, the general structure of the remaining particle stain is comparable in all cases: a dense ring-shaped stain at the contact line, generated by the well-known coffee-stain effect, but a large monolayer of particles at the center of the drop, formed by a interfacial sink flow typically generated by thermal Marangoni flow \cite{deegan2000pattern}. The presence of salts have a minor but still clearly visible effect on such central cap of particles. 

\textit{Experimental setup.} The evaporation experiments were carried out inside an incubator with temperature control (XLmulti S, Peacon) mounted on an epiflourescent inverted microscope (AxioObserver Z1, Carl Zeiss AG). A second chamber inside the incubator was used to keep constant the relative humidity around the droplet. A T-H sensor (DTH22 Adafruit, with 2--5\% accuracy) was used to monitor the relative humidity and control a flow of dry air to adjust the humidity level. In each experiment, a water droplet was gently deposited with a pipette (Eppendorf Research plus)  on a 22 mm $\times$ 22 mm, 0.2-mm-thick, glass cover slips, that was then placed inside the second chamber (Fig.~\ref{fig:sketch}).

\newcolumntype{C}[1]{>{\centering\arraybackslash}p{#1}}
\begin{table}[b]
\begin{tabular}{lC{1.48cm}C{1.48cm}C{1.48cm}C{2.05cm}}
\hline
 &UPW &ViO  &Vittel  & Gerolsteiner  \\
\hline
Na$^+$      & $<0.001$ & 15 & 8 & 12 \\
Ca$^{2+}$   & $<0.001$ & 51 & 94 & 140 \\
Mg$^{2+}$   & $<0.001$ & 5  & 20 & 49 \\
Cl$^-$      & $<0.001$ & 20 & 0 & 9\\
HCO$_3^-$   & $<0.001$ & 152& 248 & 652 \\
SO$_3^{2-}$ & $<0.001$ & 19 & 120 & 20 
\end{tabular}
\caption{Different types of water used in the experiments and respective mineral content as reported by the bottle label. Concentration values are in mg/l. }
\label{tab:waters}
\end{table}

A side-view camera was used to measure the evolution in time of the droplet profile, and therefore to calculate the evaporation rate and droplet size. The internal flow was measured  using a single-camera 3D particle tracking velocimetry (PTV) method, the General Defocusing Particle tracking (GDPT) \cite{barnkob2015general}, in combination with astigmatic optics \cite{cierpka2010calibration,rossi2014optimization}, following a procedure already used in \cite{marin2016surfactant}. The water droplets were seeded with 1-$\upmu$m-diameter fluorescent polystyrene spheres (Thermofisher) at very low concentration (less than 0.01\% w/w). Digital images of the particles in the fluid were acquired using an inverted microscope and deformed by means of a cylindrical lens placed in front of the digital camera (sCMOS, PCO GmbH). A calibration approach based on pattern recognition was used to identify the depth position of the particles from their image shape.
 
We used four types of water sources: a reference sample consisting of ultrapure water (UPW) delivered by a \mbox{Milli-Q\textsuperscript{\textregistered}} system (Direct-Q3 System, Merck) and three commercial bottled mineral waters: ViO, Vittel, and Gerolsteiner. The purity level of the ultrapure water samples was checked with conductivity measurements (18.2 M$\Omega\,$cm at 25~$^\circ$C) before each experiment. For the mineral content of the mineral waters we refer to the values given in the bottle labels and reported in Tab.~\ref{tab:waters}. Water bottled in Germany follows the directive \textit{Mineral- und Tafelwasserverordnung}, which limits the maximum contents of organic material and minerals. The declared uncertainty on the reported values is 20\% of the measured value. More detailed information about the chemical composition of bottled water in european brands can be found in reference \cite{Bertoldi:2011hh}. A first set of experiments was performed to confirm that the presence of tracer particles and the illumination source did not affect the measurements or the evaporation process (reported in the Appendices).

\textit{Results.} All experiments were carried out at an actively controlled temperature $T = 30$ $^\circ$C, relative humidity $H = 20$ \% and initial droplet volume $V = 2.5$~$\upmu$l. Immediately after deposition, the droplets assumed the classical spherical-cap shape with an average radius $a=1.28$  mm ($a_{min-max} = 1.24$--$1.34$ mm) and initial contact angle CA $=65$--$70^\circ$. During the evaporating process, the droplets maintained a pinned contact line until a CA that varied from experiment to experiment; typically it was CA $=20$--$25 ^\circ$ (around $30^\circ$ for ViO). The measurements were started at CA $=45^\circ$ and stopped when the contact line depinned.

\begin{figure}[t]
\centering
  \includegraphics[width=.45\textwidth]{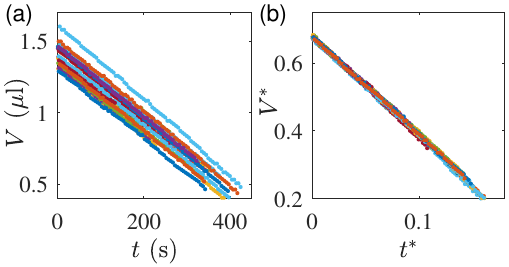}
  \caption{Evolution in time of droplets volume for all the experiments in real (a) and normalized (b) units. Time is set to 0 when the contact angle is 45$^{\circ}$. The difference in evaporation rate is given by the different radius of droplets.} 
  \label{fig:evaporation}
\end{figure}
\begin{figure*}
\centering
  \includegraphics[width=\textwidth]{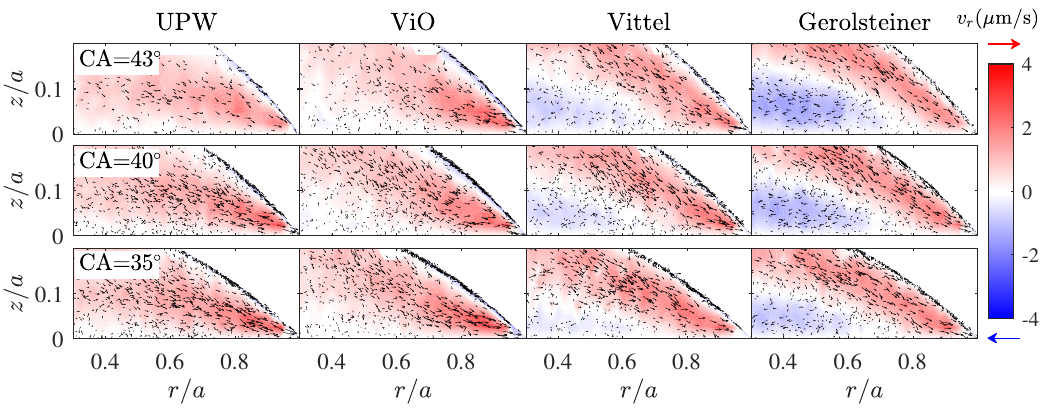}
  \caption{Internal flows for different contact angles. The colormap represent the radial velocity: positive velocity are directed outward toward the contact line. For UPW, two contribution are present: a bulk flow toward the contact line, and a interfacial flow toward the drop summit. For mineral water, a third contribution due to concentration gradient pushing the fluid in the bottom toward the drop center start to build up as the mineral content is increased (i.e. moving from ViO to Gerolsteiner).  } 
  \label{fig:velocity}
\end{figure*}

The measured droplet volume vs. time is shown in Fig.~\ref{fig:evaporation} for the full experimental ensemble. The curves are different since the drops have slightly different radii and the evaporation rate depends on the drop radius \cite{deegan1997capillary,hu2005analysis}). However all data collapse in one curve when non-dimensional units are used. We used here the classical normalization \cite{popov2005evaporative,gelderblom2011water} $V^*=V/a^3$ and $t^* = t D c_s (1-H) / a^2 \rho$, with $D$ vapor diffusion coefficient, $c_s$ saturated vapor concentration, and $\rho$ water density, all calculated at $T = 30^\circ$C. The collapse shows that the evaporation occurs following a diffusion-limited process in a very reproducible way, and that no difference is observed in evaporation rate among the different salt contents.

The velocity field is obtained summing up 3D velocity data from six different experiments for each water type. Results are reported in cylindrical coordinates and shown in Fig.~\ref{fig:velocity} (see Supplemental Material \cite{suppl} for the animated version).  
For UPW, a capillary bulk flow directed toward the contact line is observed as expected from the classical coffee-stain effect \cite{deegan2000pattern,marin2011order}. The bulk flow accelerates as the CA becomes smaller and has a maximum magnitude of about 4~$\upmu$m/s \cite{marin2011order,marin2011rush,hamamoto2011order}. The interfacial flow is clearly dominated by thermal Marangoni flow directed towards the drop summit, with a maximum magnitude of around 2~$\upmu$m/s (Fig. \ref{fig:velocity}, UPW panels). Regarding the mineral waters, the capillary flow toward the contact line is always present, however, as the content of minerals increases (i.e. ViO, Vittel, and Gerolsteiner) a stronger and stronger recirculating flow, forcing the fluid in the bottom toward the drop center, becomes visible. Similar flow patterns are normally observed in droplets containing higher concentration of chaotropic salts and has been shown to be driven by a solutal Marangoni flow which is directed in the opposite direction of the thermal Marangoni \cite{marin2018solutal}. However, our measurements show that the interfacial flow is directed towards the droplet's summit. This co-existence of \emph{a priori} contradicting flow patterns occurs more clearly for ViO and Vittel, as we will see in the next representation of the data. 

\begin{figure}
  \centering
  \includegraphics[width=0.48\textwidth]{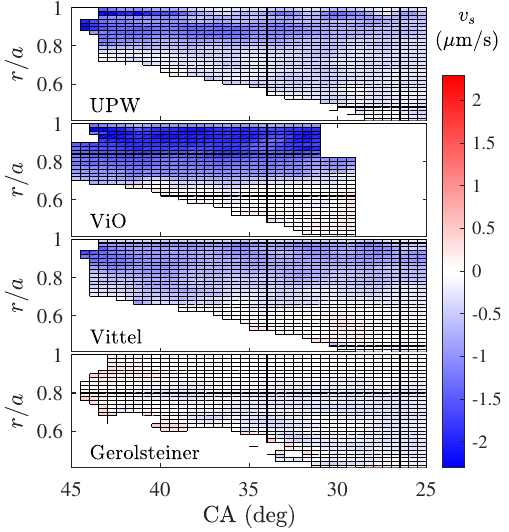}
  \caption{Interfacial flow velocity ($v_s$) as a function of the radial position and the contact angle. Negative velocities indicate flow directed toward the droplet center.}
  \label{fig:marangoni}
\end{figure}

Such a surprising effect motivates to evaluate the interfacial flows in more detail. In Fig. \ref{fig:marangoni} we plotted the drop surface velocity as a function of contact angle and normalized radial coordinate.  Note that since the depth of the PTV measurements was limited to about 0.25~mm from the bottom, the portion of surface that could be measured increases with time (i.e. as the contact angle becomes smaller).
Starting with our reference case, UPW shows a classical thermal Marangoni flow which is directed, as expected, away from the contact line (blue regions, negative radial velocity in our reference frame) and tends to decrease as the contact angle becomes smaller. This is expected since the thermal gradients should vanish as the droplet becomes thinner \cite{hu2005analysis}. The largest velocity values are located close to the contact line at about 0.8--0.9 $r/a$  (in the order of $2.5~\upmu$m/s) and decrease as we move towards the center. 
As mentioned above, mineral waters show an interfacial flow in the same direction as a thermal Marangoni flow. For ViO, its maximum strength is also around $2.5~\upmu$m/s but it survives only until a distance of about 0.7 $r/a$ where the velocity drops abruptly. 
The same behavior, just scaled with a smaller maximum flow strength of about $2.2~\upmu$m/s, is observed of Vittel. 
Interfacial flow in Gerolsteiner shows a very different trend, with significantly smaller interfacial velocities. From Fig. \ref{fig:marangoni}, we could define those regions in which the interfacial velocity is within the measurement error ($-0.2$ and 0.2 $\upmu$m/s) as \emph{stagnant regions}. In this sense, ViO and Vittel have very well defined stagnant regions at their interface for approximately $r/a>0.7$ in both cases.

\begin{figure}
\centering
  \includegraphics[width=.43\textwidth]{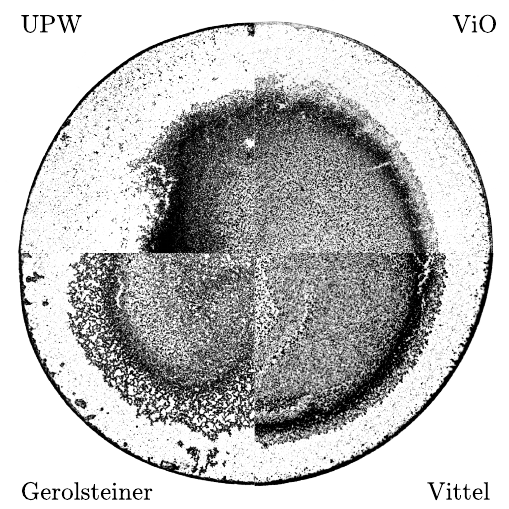}
  \caption{Final stain obtained after evaporation of the droplet for the four different types of water. The stain is formed by 1-$\upmu$m-diameter spheres dispersed in the water. In all cases the typical ring-shaped stain at the contact line can be observed. Going toward the center there is a region without particles corresponding to the portion of surface where the inwards interfacial flow is stronger. Before reaching the droplet's summit, the interfacial flow weakens, a \emph{stagnation region} appears (see Fig. \ref{fig:marangoni}), where particles tend to accumulate, specially in the case of ViO and Vittel.} 
  \label{fig:stain}
\end{figure}

Footprints of these flows can also be found in the final stains as shown in Fig.~\ref{fig:stain}. The stains are formed by the 1-$\upmu$m-diameter tracer particles contained in the droplet. A movie of the full evaporation process is included in the Supplemental Material \cite{suppl}. First, the particles are transported by the bulk flow to the contact line area. Those particles closer to the bottom substrate remain trapped in the contact line forming the classical ring-shaped stain. Those located farther away from the substrate approach the drop liquid-air interface and are then transported toward the drop's summit following the inward interfacial flow. However, these particles will typically stop when they reach the stagnant regions described above, where the interfacial flow is not strong enough to carry them farther. For ViO and Vittel, an accumulation of particles can be observed at $r/a = 0.7$, which is very likely correlated with the abrupt drop of interfacial velocity observed in Fig.~\ref{fig:marangoni} (more details are reported in the Appendices). For Gerolsteiner, the particle accumulation region starts earlier, at $r/a = 0.6$.

\textit{Discussion.} In our reference experiment with UPW, containing the minimum amount of ions and contaminants, we measure a clear and classical thermal Marangoni flow with velocity values up to 2.5 $\upmu$m/s. These values are in agreement with previous experimental observations \cite{xu2007marangoni,girard2008effect,marin2016surfactant}, however they are 2--3 order of magnitude smaller than velocities predicted by current analytical and numerical simulations \cite{hu2005analysis,girard2008effect}. This discrepancy is commonly ascribed to unavoidable contamination of insoluble surfactants in the experiments. Our results bring strong evidence of a more complex scenario. 
The experiments were carried out carefully to reduce the contamination of the samples by minimizing their exposure to the laboratory atmosphere and repeated over different weeks obtaining the same results\footnote{Note that identical results with UPW have been also obtained in different laboratories.}.
Preliminary experiments showed that tracer particles in low concentration (surfactant-free) do not affect the experiment and follow faithfully the flow. More importantly, experiments performed with UPW and mineral waters with low mineral content (notably ViO) exhibit substantially the same magnitude of the interfacial flow, which suggest that the flow observed in the mineral waters is also of  thermocapillary nature. 

On the other hand, the magnitude of the interfacial flow in mineral waters is slowed down by the presence of salts/minerals. This is expected since the presence of chaotropic salts has been shown to reverse the direction of the interfacial flow \cite{marin2018solutal}. Surprisingly, we observed the co-existence of two flow patterns: an inwards interfacial flow (restricted to a portion of the surface close to the contact line) and an additional inwards bulk flow at the bottom, closer to the glass substrate. It is also interesting to note that, in those cases in which the presence of salt is higher (see Fig. \ref{fig:stain} Gerolsteiner and Vittel), particle aggregates at the central part of the stain become more complex, ramified and with lower compactness, which are classically observed in salt-induced colloidal aggregation~\cite{carpineti1990salt}.

\textit{Conclusion.} In summary, we have shown that the presence of small amounts of chaotropic salts have a strong influence, not only on the interfacial flow of the droplet, but also in the total flow pattern inside the droplet. Our measurements identify the presence of a thermal Marangoni flow in the reference case of UPW, but we identify complex interfacial flows for diverse mineral waters. The complexity of the flow patterns observed is an evidence of the interplay between thermal and solutal Marangoni flow caused by the presence of salts. 
However, the influence of the mineral concentration in the final pattern seems to be minor. The shape and reach of the central monolayer of colloids is correlated with the interfacial flow patterns since the stagnant regions roughly coincide with the width of the particle accumulation regions.
Our results combine unprecedented observations on the interfacial flows in evaporating sessile droplets of water with small amounts of salt and show the significant sensitivity that water solutions have to ``ionic'' contamination.
To conclude, we believe that the analogy between the flow patterns in UPW and mineralized samples as ViO are a strong evidence that the divergence between experiments and numerical simulations cannot be ascribed solely to the presence of contaminants in water. In our opinion, and under the light of these results, the large discrepancy between the simulated thermal Marangoni velocities and the measured ones should be analyzed by considering a more complex thermofluidic scenario at the water-air interface \cite{Sefiane:2007hea}.

The authors acknowledge financial support by the Deutsche Forschungsgemeinschaft
KA1808/22.

\section*{APPENDIX A: 3D particle tracking} 
\label{secS:ptv}

The flow inside the droplets is measured using General Defocusing Particle Tracking (GDPT) in combination with astigmatic optics and for more detail about the measurement method and calibration principles we refer to [20, 21, 22]. The images were taken with an epifluorescence inverted microscope (AxioOberver Z1, Carl Zeiss AG), using a 10$\times$ microscope objective in combination with a cylindrical lens with focal length of 500 mm placed directly in front of the digital camera sensor (sCMOS, PCO). This optical configuration yielded  a measurement volume of $1400 \times 1300 \times 150$ $\upmu$m$^3$, with an estimated measurement uncertainty of $0.06~ \upmu$m in the in-plane direction and 1 $\upmu$m in the out-of-plane direction. The images were taken at a frame rate of 1 fps, sufficient to resolve the slow flow dynamics of the phenomena involved.  

After the GDPT evaluation, the position and velocity data collected from each experiment were converted in normalized unit. As reported in the main text, the normalization was performed following [8, 9]; 
\begin{center}
\begin{tabular}{l l c l}
Length: & $l^*$ & = & $ l/a$\\
Volume: & $V^*$ & = & $V/a^3$\\ 
Time: & $t^*$ & = & $t D c_s (1-H) / a^2 \rho$\\ 
Velocity: & $v^*$ & = & $v a \rho / D c_s (1-H)$\\
\end{tabular}
\end{center}

\noindent
with $a$ droplet radius, $D$ vapor diffusion coefficient, $c_s$ saturated vapor concentration, $H$ relative humidity, and $\rho$ water density. The droplet radius changed from experiment to experiment whereas all other parameters remained constant ($H = 20\%$ and all other parameters calculated for $T = 30^\circ$ C). Afterwards, a calibrated time-averaged smoothing was performed on each particle trajectory to reduce errors due to random noise and Brownian motion and each particle trajectory was interpolated in time to obtain position and velocity data at the desired contact angles (from 45$^\circ$ to 25$^\circ$ with steps of 0.1$^\circ$).

\begin{figure}[b!] 
\centering
  \includegraphics[width=.5\textwidth]{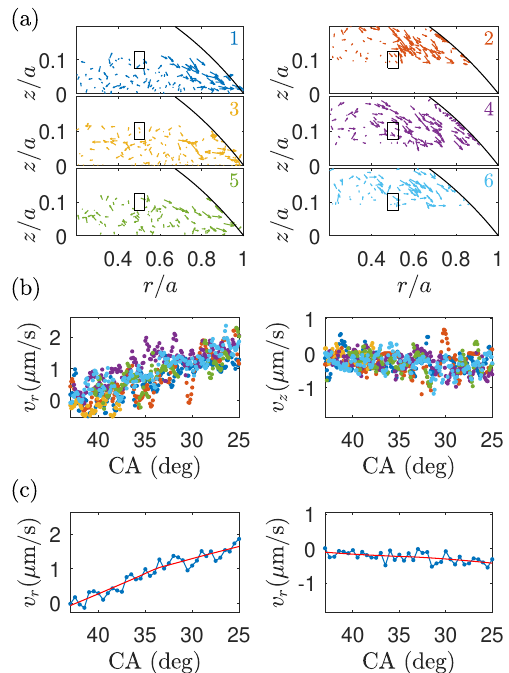}
  \caption{(a): Vector plots of the PTV measurements of six individual experiments, reported in cylindrical normalized coordinates and corresponding to a contact angle of 38$^\circ$. (b): Average velocity components inside the area indicated by the black box in the vector plots, as a function of the contact angle. (c): Final values of the velocity components in the central point of the box after combining the six measurements. The red line indicate a time-averaged value used to estimate the random error.  } 
  \label{figS:repeat}
\end{figure}

For each case, the final velocity field was obtained from six independent experiments, taken at two different height positions to cover a maximum droplet height of about 250 $\upmu$m ($z/a = 0.2$). The normalized velocity data were ensemble averaged together and interpolated on a regular grid using linear interpolation. For final presentation, the velocity data were converted in physical units using the average drop radius from the six experiments as characteristic length.

A direct estimation of the uncertainty of the final velocity values is not straightforward, due to the several post-processing steps involved, including space-time averaging and interpolation. In Fig.~\ref{figS:repeat}, we show as example (a) the six vector plots added up to create the velocity map for the droplet of Vittel water at CA = 38$^\circ$, (b) the average velocity measured inside a box around a point of coordinate $r/a = 0.5$ and $z/a = 0.1$ for the six experiments as a function of the contact angle, and (c) the final velocity data obtained for such point (blue line) after combining the six experiments. In general, the six experiments show a random fluctuation around the mean value of 0.3--0.5 $\upmu$m/s and they follow a consistent trend showing no obvious bias error. The final velocity data, obtained from an ensamble average of the six experiments, show a smaller random fluctuations that can be quantified by a comparison with a time-averaged value (red line in Fig.~\ref{figS:repeat}(c)), giving uncertainty values of 0.1--0.2 $\upmu$m/s for both the velocity components $v_r$ and $v_z$.  Similar values are consistently observed for different points (including points in top and bottom regions calculated from three experiments) and different waters.

\section*{APPENDIX B: Effect of illumination and particles on the evaporation rate} 

\begin{figure}[b!]
\includegraphics[width=.45\textwidth]{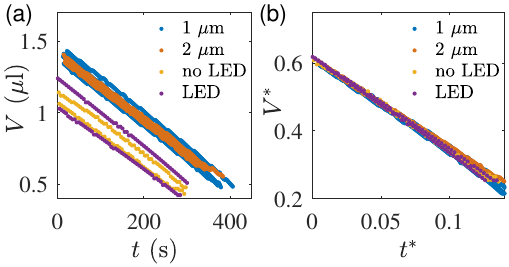}
\caption{Evolution in time of droplets volume for droplets with and without LED illumination (no particles) and with two type of particles (1-$\upmu$m-diameter particles from Thermofisher, and 2-$\upmu$m-diameter particles from Microparticles GmbH). (a) Plot in in real units, (b) plot in normalized units. Time is set to 0 when the contact angle is 45$^{\circ}$.} 
\label{figS:evaporation}
\end{figure}

A set of preliminary experiments was performed to check that the LED illumination and the presence of particles in the fluid had a negligible impact in the evaporation process and in the internal flows. 

The evaporation rate (drop volume as a function of time) of ultrapure water droplets was measured with and without the LED illumination and with and without the tracer particles. Two types of tracer particles, all made of polystyrene (PS) were used:
\begin{itemize}
\item 1-$\upmu$m-diameter PS spheres  with carboxylate as surface functional groups from Thermofisher.
\item 2-$\upmu$m-diameter PS spheres  with sulfate as surface functional groups from Microparticles GmbH. 
\end{itemize}
The particle concentration was the same used for the PTV measurements, about $ 0.01\%$ w/w. The room conditions were the same for all the experiments with $T=30$~$^\circ$C and $H = 20$~\%.

Results are presented in Fig.~\ref{figS:evaporation}. It should be noted that in our experimental setup we cannot control precisely the droplet radius and that different drop radii results in different evaporation rates as reported in Fig.~\ref{figS:evaporation}(a). However, using normalized units, the evaporation rate plots fall nicely on top of each other, proving that LED illumination and particle dispersion (at low concentration) have no measurable effect on the evaporation rate.

\section*{APPENDIX C: Reproducibility of the stain patterns}
\label{secS:rep}

\begin{figure}[b!] 
  \includegraphics[width=.5\textwidth]{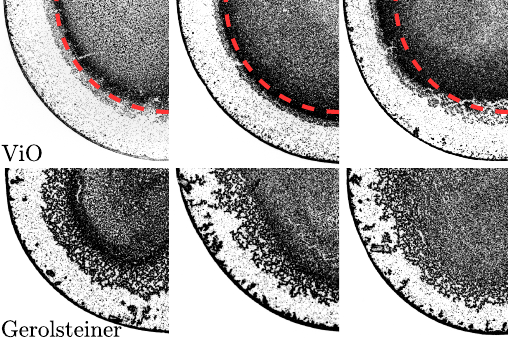}
  \caption{Stain patterns obtained from different drying experiments carried out at the same conditions. The first and second rows correspond to experiments performed with the water Vio and Gerolsteiner, respectively. The dashed line in the first row indicates the center of the stagnation region with low velocities, which coincides with a particle accumulation region.}  
  \label{figS:stains}
\end{figure}

\begin{figure*}[t!] 
\centering
  \includegraphics[width=\textwidth]{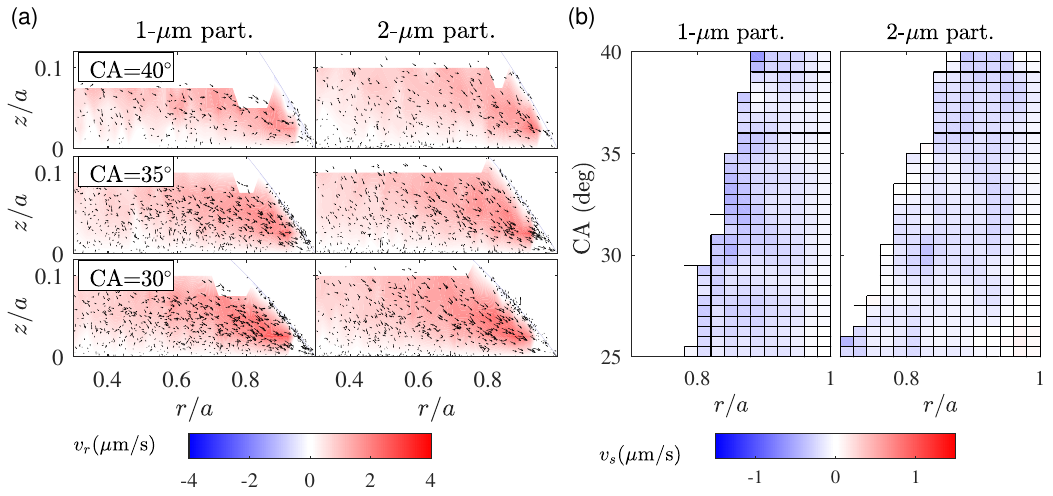}
  \caption{Comparison of flow velocities measured with 1-$\upmu$m-diameter particles and 2-$\upmu$m-diameter particles. (a) Internal flows in cylindrical coordinates for different contact angles. (b) Interfacial flow velocity as a function of the radial position and the contact angle.} 
  \label{figS:velocity}
\end{figure*}

The experimental setup allows a high reproducibility of the results as shown in the previous sections with respect to velocity measurements and evaporation rate. Consequently, also the stain left after the evaporation process is consistent and show high reproducibility. The main factor to take into account here is the de-pinning of the contact line, which must happen at approximaetly the same contact angle. Fig.~\ref{figS:stains} illustrates different drying experiments, performed with Vio and Gerolsteiner. The first picture in the row is the one reported in Fig.~5 of the main manuscript, the other pictures correspond to different experiments and performed under the same conditions. The same qualitative patterns can be observed for the same waters. The experiments have not been repeated for the other waters, but there are no reasons to expect different behaviors.

\section*{APPENDIX D: Effect of particles on the velocity measurements} 

PS spheres are used over decades as standard tracer particles for microfluidic experiments in water. In our case, the flow velocities are orders of magnitudes larger than the corresponding particle Brownian motion and the sinking velocity (polystyrene density $\rho_{PS} = $1050 kg/m$^3$ is slightly denser than water at 30 $^\circ$C), therefore we expect the particles to faithfully follow the bulk flow velocity.

Some concerns can emerge over the uncertainty of the interfacial flow measurement mainly for two reasons: The tracer particles may interact with the water-air interface, and the particles'  diameter may be too large compared to the velocity gradient in proximity of the surface (thus underestimating the actual surface velocity). To have an estimation of whether or not these issues are relevant for our case, we measured the interfacial velocity of ultrapure water droplet using two different tracer particles with different size and surface properties: The 1-$\upmu$m-diameter particles  (Thermofisher, functionalized with carboxyl groups) and the 2-$\upmu$m-diameter (Microparticles GmbH, functionalized with sulfate groups) described in the previous section. 

Results are shown in Fig. \ref{figS:velocity}. The velocity fields are obtained summing up five individual experiments for each tracer, all taken at one height position (also up to $z/a = 0.1$). Despite the differences in surface properties and particle diameters,the differences in measured velocity are, on average, 0.08 $\upmu$m/s in the bulk flow and 0.05 $\upmu$m/s in the interfacial flow, which are smaller than the estimated measurement uncertainty. The 1-$\upmu$m particles allow a sightly larger particle concentration (thus more data points and better statistics in a single experiments) and were selected as tracer particles for the main measurements. The same results are also expected for the mineral waters, since the content of minerals and particle concentration is very low to expect any different particle behavior and the main features of the flow, in terms of velocity magnitude and gradients, are similar in all experiments.

%

\end{document}